\documentstyle[12pt]{article}

\begin{document}
\begin{titlepage}
\begin{center}

March 14. 2000     \hfill    LBNL-44712 \\

\vskip .5in

{\large \bf From Einstein Nonlocality to von Neumann Reality}
\footnote{This work is supported in part by the Director, Office of Science, 
Office of High Energy and Nuclear Physics, Division of High Energy Physics, 
of the U.S. Department of Energy under Contract DE-AC03-76SF00098}

\vskip .50in
Henry P. Stapp\\
{\em Lawrence Berkeley National Laboratory\\
      University of California\\
    Berkeley, California 94720}
\end{center}

\vskip .5in

\begin{abstract}
Recent nonlocality results support a new picture of reality
built on the ideas of John von Neumann.

\end{abstract}

\end{titlepage}

\newpage
\renewcommand{\thepage}{\arabic{page}}
\setcounter{page}{1}

``Nonlocality gets more real''. This is the provocative title of a recent 
report in Physics Today [1]. Three experiments are cited. All three
confirm to high accuracy the predictions of quantum theory in experiments 
that suggest the occurrence of an instantaneous action over a large distance. 
The most spectacular of the three experiments begins with the 
production of pairs of photons in a lab in downtown Geneva. For some of
these pairs, one member is sent by optical fiber to the village 
of Bellevue, while the other is sent to the town of Bernex. 
The two towns lie more than 10 kilometers apart. Experiments on the arriving 
photons are performed in both villages at essentially the same time. What 
is found is this: The observed connections between the outcomes of these 
experiments defy explanation in terms of ordinary ideas about the nature 
of the physical world {\it on the scale of directly observable objects.}
This conclusion is announced in opening sentence of the 
Physical-Review-Letters report [2] that describes the experiment: 
``Quantum theory is nonlocal''. 

This observed effect is not just an academic matter. A possible application
of interest to the Swiss is this: The effect can be used in principle to 
transfer banking records over large distances in a secure way [3]. 
But of far greater importance to physicists is its relevance to two 
fundamental questions: What is the nature of physical reality? What is the 
form of basic physical theory? 

The answers to these questions depend crucially on the nature of physical 
causation. Isaac Newton erected his theory gravity on the idea of instant 
action at a distance. The idea was later banished from classical physics 
by Einstein's theory of relativity. However, the idea resurfaced at the 
quantum level in the debate between Einstein and Bohr. Einstein objected to 
the ``mysterious action at a distance'', but Bohr defended  ``the necessity 
of a final renunciation of the classical ideal of causality and a radical 
revision of our attitude towards the problem of physical reality''[4]. 

The essence of this radical revision was explained by Dirac at the 1927 Solvay
conference [5]. He insisted on the restriction of the application of quantum 
theory to our knowledge of a system, not to that system itself. 
This view is encapsulated in Heisenberg's famous statement [6]:

``The conception of the objective reality of the elementary particles
has thus evaporated not into the cloud of some obscure new reality
concept, but into the transparent clarity of a mathematics that represents
no longer the behaviour of the particle but rather our knowledge of this
behaviour.''  

This conception of quantum theory, espoused by Bohr, Dirac, and Heisenberg,
is called the Copenhagen interpretation. It is essentially
subjective and epistemological, because the basic reality of the
theory is `our knowledge'.

It may seem odd at first that such prominent physicists would propose this 
radical revision of the nature of physical theory. 
But they seemed to be forced to this subjective point of  view 
by certain failures of normal ideas about causation. 

In actual practice,
quantum theory often entails that an act of acquiring knowledge in one place 
instantly changes the theoretical representation of some faraway system. 
Physicists were---and are---reluctant to believe that performing a nearby 
act can instantly change a faraway physical reality. However, they know that 
``our knowledge'' of a faraway system can instantly change when we acquire  
knowledge about a nearby system, provided some properties of two systems are 
known to be strongly correlated. For example, if we know that two particles 
start at the same time  from the origin of the coordinate system, and
subsequently travel with opposite velocities, then finding one of these 
particles at the point $(x,y,z)$ allows us to `know' that the other particle 
lies, at that same instant, at the point $(-x,-y,-z)$. But we do not imagine
that the act of measuring the position of one particle {\it causes} the other 
particle to {\it be} where it is. By analogy, the instantaneous effects that 
automatically arise in quantum theory become less puzzling if one restricts 
the applications of quantum theory to ``our knowledge'', and renounces 
all efforts to understand physical reality, except to the extent that 
``physical reality'' is identified with knowledge.

This way of dodging the locality problem was attacked by Einstein, Podolsky, 
and Rosen in a famous paper [7] entitled: ``Can quantum-mechanical description 
of physical reality be considered complete?'' Einstein and his colleagues 
argue that the answer to this question is No, while Bohr argues for the 
affirmative. Given the enormity of what must exist in the universe that is 
not ``our knowledge'', it is astonishing that, in the minds of most 
physicists, Bohr has prevailed over Einstein in his claim that quantum 
theory, in a form that is explicitly restricted in application to human 
knowledge, can be considered to be a complete description of physical reality.
This majority opinion stems, I believe, more from the lack of a promising 
alternative candidate than from any decisive logical argument.

Einstein, commenting on the Copenhagen position, said:  ``What I dislike 
about this kind of argument is the basic positivistic attitude,  which from 
my view is untenable, and seems to me to come to the same thing as Berkeley's 
principle, {\it esse est percipi} [9]. Many other scientists agree.
For example, Murray  Gell-Mann [10] asserts: ``Niels  Bohr brainwashed
a whole generation into believing that the problem was solved 
fifty years ago''. Gell-mann has been pursuing with James Hartle [11]
an ambitious program, built on ideas of Everett [12] and of Griffith [13], 
that aims to construct a quantum theory that is more complete than the
Copenhagen version. This effort, and others like it, are fueled by the 
opinion that to integrate quantum theory into cosmology, and to understand 
the evolutionary process that has  produced creatures that can have knowledge 
akin to ``our knowledge'', one needs to have a theory of the evolving reality 
in which those creatures are imbedded.

It is in this context of such efforts to construct a more complete theory 
that the significance of the quantum nonlocality experiments lies.
The point is this: If nature really is nonlocal, as these experiments suggest,
then there is a simple theory of reality that encompasses both our knowledge 
and a real objective physical world in which that knowledge is embedded. It
describes also the dynamical connection between these two aspects of reality. 
This theory is obtained by combining relativistic quantum field theory with 
the version of quantum theory developed by John von Neumann [14].

All physical theories are, of course, provisional, and subject to future 
revision or elaboration. But at a given stage in the development of science 
the contending theories can be evaluated on many grounds, such as utility, 
parsimony, predictive power, explanatory power, conceptual simplicity, 
logical coherence, and aesthetic beauty. The relativistic version of 
von Neumann's theory fares well on all of these counts.
 
The essential difference between von Neumann quantum theory 
and Copenhagen quantum theory lies in the way measuring devices are 
treated. In the Copenhagen approach, the measuring devices are excluded 
from the world described in the mathematical language of quantum 
theory. The measuring device are described, instead, by ``the same means 
of communication as the one used in classical physics'' [15]. This approach 
renders the theory pragmatically useful but physically obscure. It links 
the theory to ``our knowledge'' of the measuring devices in a useful way, 
but upsets the unity of the physical world. This tearing asunder of the 
physical world creates huge theoretical problems, which are ducked in the 
Copenhagen approach by renouncing man's ability to understand 
reality.  

The mathematical rules of quantum theory specify clearly how the measuring 
devices are to be included in the quantum mechanically described physical 
world. Von Neumann first formulates rigorously the mathematical structure 
that quantum phenomena seem to force upon us, and then follows where that 
mathematics leads. It leads first to the incorporation of the measuring 
devices into the quantum mechanically described physical universe, and 
eventually to the inclusion of {\it everything} built out of atoms and their 
constiuents. Our bodies and brains thus become, in von Neumann's approach, 
parts of the quantum mechanically described physical universe. Treating the 
entire physical universe in this unified way provides a conceptually simple 
and logically coherent theoretical foundation. The Copenhagen alternative 
of leaving out of this description parts of the physical universe that are 
interacting with the parts retained severely disrupts the logical 
coherence of the theoretical structure.

Copenhagen quantum theory claims to be complete. That claim stems from 
the fact that all of validated predictions of classical physical theory can 
deduced from it, together with a host of quantum predictions, many validated,
and none known to fail. Bohr argues that all possible predictions 
pertaining to connections between outcomes of human observations of the 
devices that probe atomic systems are obtainable from Copenhagen quantum 
theory.

Von Neumann quantum theory encompasses, in  principle, all of the prediction 
of Copenhagen quantum theory. It postulates, for each observer, that each 
increment in his knowledge is connected to a corresponding `reduction' of the 
state of his brain: the new reduced state is obtained from the old state by 
eliminating all parts that are incompatible with his new knowledge. This rule 
is a direct application, at the level of the brain of the observer, of the 
rule that Copenhagen quantum theory applies at the level of the measuring 
device, and the equivalence is the two formulations arises from the causal
connection that is needed to effect a good observation. 

But von Neumann quantum theory gives, in principle, much  more than Copenhagen
quantum theory can. By providing an objective description 
of the entire history of the universe, rather than merely rules connecting 
human observations, von Neumann's theory provides a quantum framework for 
cosmological and biological evolution. And by including the body and brain of 
the observer as well as his knowledge, and also the dynamical laws that 
connects these two realities, the theory provides a coherent 
framework for understanding the relationship between mind and matter [16].  

Von Neumann's rules are, of course, expressed in neat mathematical 
form. [See Box 1]

------------------------------------------------------------------

{\bf Box 1: von Neumann Quantum Theory}

The evolving state of the universe is represented by an operator S(t).

The state of any subsystem, b, is represented by 
$$
S_b(t)= Tr_b S(t),
$$
where $Tr_b$ stands for the partial trace over all variables other than those
that define the subsystem b.

The system S(t) evolves between reductions via the equation 
$$
S(t+ \Delta t)= \exp (-iH\Delta t)S(t)\exp (+iH\Delta t),
$$
where H is the energy operator. Each reduction is associated with a
quantum information processor b and a projection operator P 
that acts like the identity on all degrees of freedom other than those 
that define b. The reduction proceeds in two steps. First a question is posed
by the processor. This is represented by the von Neumann process I:
if $S(t-0)$ is the limit of $S(t')$ as $t'$ approaches $t$ from below then
$$
S(t) = P S(t-0)P + (1-P) S(t-0) (1-P). 
$$
Then nature chooses an answer, $P=1$ or $P=0$, according to
the rule

$S(t+0) = PS(t)P$ with probability $Tr PS(t)/Tr S(t)$,

or

$S(t+0) = (1-P) S(t) (1-P)$ otherwise.

-------------------------------------------------------------------

{\bf Reconciliation with Relativity}

von Neumann quantum theory gives a logically simple  mathematical 
description of an evolving physical world that is linked to human 
experiences by specified dynamical equations. But there is one major
problem: reconciliation with the theory of relativity. This problem arises 
from the fact that von Neumann formulated his theory in the nonrelativistic 
approximation.

The problem has two parts. The first is resolved by simply replacing 
the nonrelativistic theory used by von Neumann with relativistic 
quantum field theory. To deal with the link to human knowledge, 
and hence to the predictions of Copenhagen quantum theory, one needs to 
consider human brains. Quantum electrodymamics is the relevant field theory, 
and the pertinent energy range is that of atomic and molecular interactions. 
I shall assume that whatever high-energy theory eventually prevails 
in quantum physics, it will reduce to quantum electrodynamics in this regime.

The second problem is this: von Neumann's theory is built on the Newtonian
concept of the instants of time, `now', each of which extends over all space. 
The evolving state of the universe S(t) is defined to be the state of the 
entire universe at the instant of time t. The formulations of relativistic 
quantum field theories by Tomonaga [17] and Schwinger [18] have corresponding 
spacelike surfaces $\sigma $. As Pauli once strongly emphasized to me, these 
surfaces, while they may give a certain aura of relativistic invariance, do 
not differ significantly from the constant-time surfaces that appear in
the nonrelativistic approximation. Indeed, all efforts to eliminate from 
quantum theory this preferred status of time have  proved futile. 

To obtain a relativistic version of von Neumann's theory one needs to 
identify von Neumann's constant-time surfaces  with certain 
special spacelike surfaces $\sigma $ of the formulations of Tomonaga 
and Schwinger. To achieve an {\it objective} quantum theory of reality 
theory, these preferred instants ${\it now }$ must be objective features 
of nature.

Giving special physical status to a particular sequence of spacekike surfaces 
runs counter to certain ideas spawned  by the  theory of relativity. However, 
the astronomical data [19] indicates that there is a preferred sequence of
`nows' that define spacelike surfaces in which, for the early universe,
matter was distributed almost uniformly in density, mean local velocity, and
temperature. 

I shall assume that there is a preferred advancing sequence of spatial 
surfaces, and that in the early universe these surfaces are defined by the 
cosmologically preferred frame.

{\bf Nonlocality and Relativity}

This theory immediately accounts for the faster-than-light transfer of 
information that seems to be entailed by the nonlocality experiments: 
the reduction of the state S(t) of the universe on the occasion of the 
earlier of the two measurement, which (in the cited experiment) occurs in one 
of the two villages, has, according to this theory, an immediate effect on 
the evolving state S(t) of the universe, and hence an immediate effect also 
on the propensities for the various possible outcomes of the measurement 
performed slightly later in the other village. 

Such an instantaneous transfer of information is widely held to be impossible:
it is believed to violate the precepts of the theory of relativity. 
But does it?

The theory of relativity was originally formulated within classical physical
theory, and, in particular, for a deterministic theory. In that case the 
entire history of the universe could be conceived to be laid out for all 
times in a four-dimensional spacetime. The idea of ``becoming'',
or of the gradual unfolding of reality, has no natural place in
this deterministic conception of the universe. 

Quantum theory is a different kind of theory: it is formulated
as an indeterministic theory. Determinism is relaxed in two
ways. First of all, freedom is granted to experimenters to chose
which measurements they will perform. Second,  Nature is then required
to choose the outcome of any experiment that is actually performed, 
subject to statistical conditions. Nature is not required to choose
an outcome for a contemplated alternative possible experiment that is 
not actually performed.

In view of these deep structural differences there is a question of 
principle regarding how the idea of no faster-than-light transfer of 
information should be carried over from the deterministic classical 
idealization to the indeterministic quantum reality.

Relativistic quantum field theory is the canonical relativistic
generalization of nonrelativistic quantum theory. That theory has two key
relativistic properties: (1), All of its predictions about outcomes of
measurements are independent of the coordinate frame used to define the 
advancing sequence of constant-time surfaces `now'; and (2), No  {\it signal} 
can be transmitted faster than light. [A ``signal'' is a 
{\it controllable} transfer of information: it is a transfer that allows 
a sequence of bits composed by a sender to be conveyed to a receiver.]
However, the theory explicity exhibits other transfers of information
that do not conform to the no-faster-than-light rule. These transfers
are associated with the reduction events. Within the theory these transfers
act instantaneously along the spatial slices of Tomonaga and Schwinger, once 
this sequence of advancing constant-time surfaces $\sigma $ is fixed.
The locus of these transfers can be shifted by shifting these surfaces,
but, within the theory, these transfers cannot be eliminated.

As mentioned above,
the usual way of dealing with these explicit faster-than-light-transfers
in relativistic quantum field theory is to say that their appearance shows 
that the theory cannot be interpreted realistically: the theory {\it must}
be about ``our knowledge'', as Bohr and company claim, rather that about 
reality itself. There is no puzzle about the fact that our knowledge 
about a faraway system can suddenly change when we acquire  here information 
about some system that is strongly correlated with that far away system.
But it is maintained that reality itself cannot behave in this way.

That is indeed the widely held prejudice. But there is no theoretical
or empirical evidence that supports it. Indeed, both theory and the 
nonlocality experiments appear to contradict it. It is thus a metaphysical
prejudice with no scientific basis. Scientific theories should be judged on 
the basis of the criteria mentioned above, rather than on the basis of a
pure metaphysical prejudice. Renouncing our ability to understand the 
world around us is a price too heavy to pay to preserve a mere prejudice. 

If that metaphysical prejudice is wrong, then all of the contortions  
and evasions and renunciations that characterize the subjective 
Copenhagen interpretation can be discarded: one can reaffirm, with  
Einstein, the traditional idea that we should pursue, without self-imposed 
limitations, our efforts to understand the world around us and our connection 
to it: if that  metaphysical prejudice is wrong then we can return to the idea 
that the proper goal of science is to understand the objectively existing 
reality of which `our knowledge' is a tiny part.

{\bf Is Nonlocality Real?}

The claim that von Neumann's theory can describe objective reality rests 
heavily on the assumption that nonlocality is real. But how strong is the 
evidence for this? Is there really any credible evidence that information is
transferred over spacelike intervals?

The evidence is very strong that the predictions  of quantum theory
are valid in these experiments involving pairs of measurementss
performed at essentially the same time in regions lying far apart.
But the question is this: Can we validly argue from the empirically 
supported premise that these predictions of quantum theory are correct to 
the conclusion that nature must transfer information over spacelike
intervals?

The usual arguments for nonlocality stem from the work of John Bell [20]. 
Pondering the issue debated by Einstein and Bohr, Bell proposed the following
approach: Assume that quantum theory is indeed incomplete, as Einstein claimed,
and hence that there are variables other than those that appear in Copenhagen
quantum theory. Then formulate a locality requirement in terms of these
extra variables, which are then called ``local hidden variables'', and prove 
that the existence of such variables is incompatible with the  assumed
validity of certain predictions of quantum theory. 

Bell  was able to prove such a contradiction. This proof showed,
basically, that Einstein was wrong: Einstein's asssumption that quantum theory
is both incomplete and local is not viable.

But that sort of argument is no  proof, or even indication, that quantum 
theory is nonlocal. The more plausible conclusion, for quantum physicists, 
is that nature is local, but that Einstein was wrong in claiming that 
quantum theory is incomplete: the hidden-variable assumption is wrong. 

{\bf Eliminating Hidden Variables}

The argument of Bell is essentially different from that of Einstein,
Podolsky, and Rosen. The former shows the ideas of Einstein cannot
all be correct; the  latter aimed to show that the ideas of Bohr cannot
all be correct.

The problem faced by Einstein and his colleagues was to mount 
{\it within the quantum framework} an argument that involved a consideration 
of possible outcomes of {\it alternative} possible experiments. The
difficulty is that quantum philosophy explicity rejects the notion that the 
outcomes of two alternative possible measurements are both physically well 
defined. Indeed, that limitation was precisely the idea that Einstein and 
company wanted to challenge. But then they had to mount an argument that dealt
with alternative possible measurements without contravening {\it in their 
premises} the very precept that they were challenging. 

Their strategy was to introduce the outcomes of alternative possible 
experiments via a locality requirement on physical reality that seemed 
undeniable. The strategy succeeded: Bohr was forced into a very fragile
position that depended upon entangling physical reality with predictions,
and hence with knowledge:

``{\it ...an influence on the very conditions which define the possible
types of predictions regarding future behavior of the system.} Since these
conditions constitute an inherent element of any phenomena to which the term
`physically reality' can be properly attached we see that the argument of 
mentioned authors does not justify their conclusion that quantum-mechanical
description is essentially incomplete.''[8]

I shall pursue here a strategy similar to that of Einstein and his colleagues,
and will be led to a conclusion similar to Bohr's, namely that quantum 
physical reality is entwined with knowledge, and does involve some subtle sort
of nonlocal influence.

I introduce alternative possibilities by combining two ideas
embraced by Copenhagen philosophy. The first of these is the freedom of 
experimenters to choose which measurements they will perform. In Bohr's words:

``The freedom of experimentation, presupposed in classical physics,
is of course retained and corresponds to the free choice of experimental
arrangements for which the mathematical structure of the quantum
mechanical formalism offers the appropriate latitude.''[15]

This assumption lies at the foundation for Bohr's notion of complementarity:
some information about all the possible choices is simultaneously present
in the quantum state, and Bohr wants to provide the possibility that
any one of the mutually exclusive alternatives might be used. No matter
which choice the experimenter makes, the associated set of predictions
is supposed to hold.

The second idea is the condition of no backward-in-time
causation. According to quantum thinking, experimenters are to be considered 
free to choose which measurement they will perform. Moreover, if an outcome 
of a measurement appears to an observer at a time earlier than some time $T$, 
then this outcome can be considered to have been fixed independently of 
which experiment will be {\it freely chosen} and performed by another 
experimenter at times later than $T$: the later choice is allowed go either way
without disturbing the outcome that has already appeared to observers earlier.
For whichever choice is eventually made at the later time, the relevant 
prediction of quantum theory is supposed hold. This no-backward-in-time 
influence condition is assumed to hold for at least one coordinate system 
(x,y,z,t).

These two conditions are, I believe, compatible with quantum thinking. They
contradict no quantum precept or combination of  quantum predictions. 
They, by themselves, lead to no contradiction. But they do involve
the contemplation of alternative possibilities, and provide the needed 
logical toe-hold.

{\bf The Hardy Experimental Setup}

To get a nonlocality conclusion like the one obtained from  Bell-type
theorems,  but without contravening the precepts  of quantum theory,
it is easiest to consider an experiment of the kind first  discussed
by Lucien Hardy [21]. The setup is basically similar to the ones considered 
in proofs of Bell's theorem. There are two spacetime regions, L and R, that 
are ``spacelike separated''. This condition means that the two regions are 
situated far apart in space relative to their extensions in time, so that 
no point in either region can be reached from any point in the other 
without moving either faster than the speed of light or backward in time.
This means also that in some frame, which I take to be the coordinate 
system (x,y,z,t) mentioned above, the region L lies at times greater  
than time $T$, and region  R lies earlier than time $T$.   

In each region an experimenter freely chooses between two possible
experiments. Each experiment will, if chosen, be performed within that region,
and its outcomes will appear to observers within that region.
Thus neither choice can affect anything in the other region without 
there being some influence that acts over a space-like interval.

The argument involves four predictions made by quantum theory
under the Hardy conditions. These conditions are described in Box 2.

--------------------------------------------------------------------

{\bf Box 2: Predictions of quantum theory for the Hardy experiment.}

The two possible experiments in region  L are labelled L1 and L2.

The two possible experiments in region  R are labelled R1 and R2.

The two possible outcomes of L1 are labelled L1+ and L1-, etc.

The Hardy setup involves a laser down-conversion source that emits a pair 
of correlated photons. The experimental conditions are such that 
quantum theory makes four (pertinent) predictions:\\ \\
1. If (L1,R2) is performed and L1- appears in L then R2+ must appear in R.\\
2. If (L2,R2) is performed and R2+ appears in R then L2+ must appear in L.\\
3. If (L2,R1) is performed and L2+ appears in L then R1- must appear in R.\\
4. If (L1,R1) is performed and L1- appears in L then R1+ appears sometimes
   in R.\\ 

The three words ``must'' mean that the specified outcome is predicted
to occur with certainty (i.e., probability unity).\\ 
---------------------------------------------------------------------------

{\bf Two Simple Conclusions}

It is easy to deduce from our assumptions two simple conclusions.

Recall that region R lies earlier than time $T$, and that region L lies
later than time $T$.

Suppose the actually selected pair of experiments is (R2, L1), and
that the outcome L1- appears in region L. Then prediction 1 of 
quantum theory entails that R2+ must have already appeared in R prior to time
$T$. The no-backward-in-time-influence condition then entails that this 
outcome R2+ was fixed and settled prior to time $T$, independently of 
which way the later free choice in L will eventually go: the outcome in region
R at the earlier time would still be R2+ even if the later free choice 
had gone the other way, and L2 had been chosen {\it instead of} L1.

Under this alternative condition (L2,R2,R2+) the experiment L1 is not 
performed, and there is no physical reality corresponding to its outcome. 
But in this alternative case L2 is performed, and hence L2 must have an 
outcome. Prediction 2 of quantum theory asserts that it must be  L2+. This 
yields the following conclusion: 

Assertion A(R2):

If (R2,L1) is performed and outcome L1- appears in region L, then if
the choice in L had gone the other way, and L2, instead of L1, had been
performed in L then outcome L2+ would have appeared there.  

Because we have two predictions that hold with certainty, and the two
strong assumptions of `free choice' and `no backward causation', it is 
not surprising that we have been able to derive this conclusion. In an 
essentially deterministic context we are often able to deduce from
the outcome of one measurement what would have happened if we had made,
instead, another measurement. If the actual outcome has a unique
{\it precondition}, which leads to a unique outcome of the alternative
possible measurement then we can draw a conclusion of this kind.

Consider next the same assertion, but with R2 replaced by R1:

Assertion A(R1):

If (R1,L1) is performed and outcome L1- appears in region L, then if 
the choice in L had gone the other way, and L2, instead of L1, had been
performed in L then outcome L2+ would have appeared there.  

This assertion cannot be true. The fourth prediction of quantum theory asserts 
that under the specified conditions, L1- and R1, the outcome R1+ appears
sometimes in R. The no backward-in-time-influence condition ensures that this
earlier fact would not be altered if the later choice in region L had  been L2.
But A(R1) asserts that under this altered condition L2+ would appear in  L.
The third prediction then entails that R1- must always appear in R.
But that contradicts the earlier assertion that R1+ sometimes appears in R.

The fact that A(R2) is true and A(R1) is false means  that the choice
made in region R between R2 and R1 converts from necessarily true to 
necessarily false a statement whose truth or falsity is determined wholly 
by connections between possible events located in a region L that is spacelike
separated from the region R where the choice between R2 and R1 is made.
This is a theoretical constraint on any model that satisfies the assumptions
of the proof. It means that any model that satisfies these assumptions
must have some way of transferring information from region R to region L.

Stated more physically, our assumptions entail the existence of a constraint
connecting the outcomes that nature can choose in region L under the different
conditions that the experimenter in region L can choose to set up there,
and this constraint takes one or the other of two incompatible forms 
depending on whether the experimenter in region R chooses to perform 
experiment R1 or R2. 

It can be concluded that any model of nature that conforms to the 
predictions and general precepts of quantum theory embodied in our assumptions 
must accomodate tranfers of information over spacelike intervals. Hence the 
presence of such transfers in a putative objective theory of reality not only 
does not disqualify the theory, but constitutes, rather, a necessary property:
any theory of reality that satisfies the assumptions of the proof
must provide for tranfers of information of the kind demanded by the proof.

The physical basis of the argument is the set four predictions of quantum
theory. Although the {\it derivation} of these predictions involves quantum 
entities, such as photons, it is only the predictions themselves, not their
derivation, that enter into the argument. These predictions are about 
large-scale experiments and large scale outcomes. Thus the argument itself 
is expressed completely in terms of big things. It shows that certain 
classical ideas about causation cannot be maintained for big objects 
separated by large distances.

{\bf The World as Knowledge}

The objective quantum theory discussed here rejects the Copenhagen 
renunciation of our ability to understand the objective sources of our 
knowledge. But it accepts many other Copenhagen precepts.  
It conforms to Bohr's claim of ``the necessity of a final renunciation of the 
classical ideal of causality and a radical revision of our attitude towards 
the problem of physical reality'': the nonlocal transfers violate the 
``the classical ideal of causality'', and the conception of physical reality
is radically altered.

To appreciate the nature of this alteration note that the key physical process 
involves the posing and answering of question with two allowed answers, 
$P=1$ and $P=0$. Thus the basic dynamical process is {\it informational} in 
character. The dynamics involves two kinds of choices that are
indeterministic, at the present stage of theoretical development. Each 
quantum processor chooses questions, and nature chooses the answers. These 
answers are stored in the evolving physical reality S(t). This stored 
compendium of discrete answers has causal power: S(t) specifies the 
propensities for the posing and answering of future questions. Once the 
physical world is understood as a stored compendium of locally efficacious 
bits of knowledge, the instant transfers can be understood in terms of 
changes in ``knowledge''. 

In the Copenhagen interpretation the pertinent knowledge was ``our knowledge'':
it was conscious human knowledge of the kind we can describe and 
communicate to other human beings.  This knowledge is the foundation
of human science. Von Neumann was concerned with this kind of 
knowledge, because he needed to show that his theory could generate the 
predictions of Copenhagen quantum theory. Although human beings, and human 
knowledge, play, therefore, a special role in the theory in its  present state
of historical development, our species should play no special role in a truly 
objective description of nature.  Von Neumann's theory has, accordingly,  
been formulated  here in terms of the more general concept of quantum 
information processors, of which human beings are the paradigmatic examples. 
However, other creatures and physical systems cannnot excluded, a priori, and 
the concept of ``knowledge'' will eventually need to be developed [9] to 
accommodate the precursors of human knowledge, namely the more primitive 
forms from which  human knowledge emerged.

\begin{center}
\vspace{.2in}
\noindent {\bf References}
\vspace{.2in}
\end{center}

1. Physics Today, December 1998, p. 9.

2. W. Tittle, J. Brendel, H. Zbinden, and N. Gisin, 
        Phys. Rev. Lett. {\bf 81}, 3563 (1998).

3. W. Tittle, J. Brendel, H. Zbinden, and N. Gisin, 
        Phys. Rev. {\bf A59}, 4150 (1999).

4. N. Bohr, Phys. Rev. {\bf 48}, 696 (1935).

5. P.A.M. Dirac, at 1927 Solvay Conference {\it Electrons et photons:
Rapports et Discussions du cinquieme Conseil de Physique}, Gauthier-Villars, 
Paris, 1928.

6. W. Heisenberg, Daedalus {\bf 87}, 95-108 (1958).

7. A. Einstein, N. Rosen, and B. Podolsky, Phys. Rev. {\bf 47}, 777 (1935). 

8. N. Bohr, Phys. Rev. {\bf 48}, 696 (1935).

9. A. Einstein, in {\it Albert Einstein: Philosopher-Physicist},
   ed, P. A.  Schilpp,   Tudor, New York, 1951.  p.669.

10. M. Gell-Mann, in {\it The Nature of the Physical Universe: the 1976
    Nobel Conference}, Wiley, New York, 1979, p. 29. 

11. M. Gell-Mann and J. Hartle, in {\it Procedings of the 3rd International
Symposium on the Foundations of Quantum theory in the Light of New Technology},
eds. S Kobayashi, H. Ezawa, Y. Murayama, and S. Nomura, Physical Society
of Japan, Tokyo, 1990.

12. H. Everett III. Reviews of Modern Physics, {\bf 29}, 454 (1957). 

13. R. B. Griffiths, J. Stat. Mech. {\bf 36}, 219 (1984).

14. J. von Neumann, {\it Mathematical Foundations of Quantum Mechanics},\\ 
    Princeton University  Press, Princeton, NJ, 1955;\\
    Translation from the 1932 German original. 

15. N. Bohr, Atomic Physics and Human Knowledge, Wiley, New York,
    1958, p.88, p.72.   

16. H. Stapp, ``Attention, Intention, and Will in Quantum Physics''\\
in Journal of Consciousness Studies, {\bf 6}, 143 (1999);\\ 
and ``Quantum Ontology and Mind-Matter Synthesis in\\
{\it Quantum Future: from Volta and Como to the Present and Beyond},\\
eds. Ph. Blanchard and A. Jadczyk, Springer, Berlin, 1999, p.156;\\
``Decoherence, Quantum Zeno Effect, and the Efficacy of Mental Effort:\\
Closing the Gap Between Being and Knowing.'',\\
Lawrence Berkeley National Laboratory Report LBLN-45229.

17. S. Tomonaga, Progress of Theoretical Physics, {\bf 1},  (1946) 

18. J. Schwinger, Physical Review, {\bf 82}, 914 (1951).

19. G.F. Smoot et. al., Astrophysical Journal {\bf 396}, L1 (1992).

20. J.S. Bell, Physics, {\bf 1}, 195 (1964); and in {\it Speakable and
    Unspeakable in Quantum Mechanics.} Cambridge Univ. Press, (1987) Ch. 4;   
    J. Clauser and A. Shimony, Rep. Prog. Phys. {\bf 41}, 1881 (1978).

21. L. Hardy, Phys. Rev. Lett. {\bf 71}, 1665 (1993):\\
    A. White, D. F. V. James, P. Eberhard, and P.G. Kwiat,\\
    Physical Review Letters, {\bf 83},  3103  (1999).

\end{document}